\documentclass[11pt]{article}
\usepackage{jheppub}
\usepackage{tensor}
\usepackage{cleveref}
\usepackage{framed}
\usepackage{amsmath,amssymb,amsfonts,graphicx,tensor}
\usepackage{palatino}
\usepackage{newpxmath}
\usepackage{bbold}
\usepackage{subcaption}
\usepackage{graphicx}
\usepackage{enumitem}
\usepackage{braket}
\usepackage[dvipsnames]{xcolor}
\usepackage{comment}
\usepackage{caption}

\newcommand{\tr}{{\rm tr}}

\usepackage{amsthm,mathtools}

\newcommand{\id}{\mathrm{id}}

\usepackage{bbm}
\usepackage{float}
\usepackage{tcolorbox}

\title{Algebras for generalized entanglement wedges}
\author{Abhisek Sahu, Jeremy van der Heijden, Mark Van Raamsdonk, Rana Zibakhsh}
\emailAdd{abhi@phas.ubc.ca}
\emailAdd{jeremy.vanderheijden@ubc.ca}
\emailAdd{mav@phas.ubc.ca}
\emailAdd{rzibakhsh@phas.ubc.ca}
\affiliation{Department of Physics and Astronomy, University of British Columbia,\\
6224 Agricultural Road, Vancouver, B.C., V6T 1Z1, Canada}

\date{}

\abstract{In asymptotically AdS spacetimes, the mathematical structure of the set of entanglement wedges reflects the algebraic structure of the underlying holographic description. For more general spacetimes, Bousso and Penington  (BP) have recently proposed a generalization of entanglement wedges sharing many of the same properties as usual entanglement wedges. In this paper, we explore the hypothesis that each generalized entanglement wedge can be associated with an algebra in the (generally unknown) fundamental description (in a semiclassical limit). We postulate features of the map from entanglement wedges to algebras that provide a natural algebraic interpretation for some of the basic mathematical properties of the set of entanglement wedges. Quantitatively, we suggest a possible generalization of the Ryu-Takayanagi formula that associates the gravitational entropy of a generalized entanglement wedge with an entropic quantity for the associated algebra. Through this assignment, inclusion monotonicity and strong-subadditivity properties shown by BP for generalized entanglement wedges would follow from various inequalities satisfied by algebraic entropies. We include a detailed appendix reviewing relevant algebraic background, including a discussion of algebraic entropies and their inequalities.}

\begin{document}

\maketitle
\section{Introduction}

What are the mathematical structures that underlie the description of quantum gravity for general spacetimes? 

In quantum gravity systems defined using standard holography, the fundamental description is via quantum mechanics or quantum field theory, so the mathematical framework is familiar. We have some underlying Hilbert space, a state that encodes some specific (quantum) spacetime, and an algebra of operators corresponding to observables of the theory. But for more general spacetimes such as the cosmological spacetimes relevant to our universe, we are ignorant even of the basic mathematical structure on which a fundamental description would be based. 

In \cite{VanRaamsdonk:2009ar}, one of us suggested that a hint to the general structure of quantum gravity might be the fact that in ordinary examples of holography, we 
have a nested structure in which certain subsystems of the holographic theory define complete quantum systems in their own right, and these quantum subsystems are dual to certain subsystems of the gravity theory now known as entanglement wedges \cite{Czech_GravityDual_2012, Wall:2012uf, HeadrickHubenyLawrenceRangamani2014,DongHarlowWall2016}. The geometrical and causal structure of entanglement wedges directly reflects the mathematical structure of the underlying quantum theory, with each entanglement wedge corresponding to a specific quantum subsystem and the properties of entanglement wedges (e.g. inclusion, spacelike separation, generalized entropy) relating directly to properties of the associated quantum subsystems (inclusion, commutativity, subsystem von Neumann entropies).

Recently, Bousso and Penington have proposed a class of regions in general spacetimes that share many of the key features of usual entanglement wedges in asymptotically AdS spacetimes \cite{Bousso:2022hlz, Bousso:2023sya}. In this paper we explore the idea that these generalized entanglement wedges in general spacetimes might also correspond to some quantum subsystems in the (generally unknown) fundamental description. For ordinary entanglement wedges, the dual subsystems are spatial subsystems of the dual quantum field theory. We expect that the generalized entanglement wedges correspond to a more general type of quantum subsystem, that of a subalgebra of observables together with a state that assigns a value to each of these.

Thus, we will explore the hypothesis that for general gravitational theories in a semiclassical limit where spacetimes regions are well-defined, there is a map\footnote{This is similar to and inspired by the subregion-subalgebra duality of Leutheusser and Liu \cite{Leutheusser:2022bgi}. 
We will discuss the relation to this work below.}
\begin{equation}
W \to (\mathcal{A}_W, \omega_W) \; ,
\end{equation}
between generalized entanglement wedges and corresponding algebras/states in the underlying quantum description.\footnote{In this paper, we will only consider the situation where the entanglement wedges correspond to subsets of a spatial slice fixed by time-reflection in a static or time-reflection symmetric spacetime. We leave an investigation of the correspondence for more general entanglement wedges for future work.}

Assuming such a correspondence,
we will postulate features of this map that allow us to assign an algebraic meaning/origin to all of the the following mathematical features of the set of generalized entanglement wedges:
\begin{enumerate}
    \item There is an inclusion relation that defines a partial ordering of the wedges and an empty wedge that is included in all wedges.
    \item There is a symmetric binary relation on pairs of BP wedges that tells us whether the BP wedges are spacelike separated or not.
    \item There is an intersection operation $\{W_i\} \to \wedge_i W_i$ that allows us to assign a BP wedge to any collection of BP wedges (this is the usual geometrical intersection). 
    \item There is a join operation $\{W_i\} \to \vee_i W_i$ that allows us to assign a BP wedge to any collection of BP wedges (this is the intersection of all BP wedges that include $\cup_i W_i$ or the BP wedge with the smallest generalized entropy that includes $\cup_i W_i$). 
    \item The wedges may be placed in equivalence classes (roughly, equivalent wedges approach the conformal boundary in the same way) such that to each pair of BP wedges $W_1, W_2$ in a class, we have an antisymmetric function $\Delta S_{gen}(W_1, W_2)$ that is finite and positive whenever $W_2 \subset W_1$. 
    \item For wedges in the class of the empty wedge, we have a function $S_{gen}(W)$ (the generalized entropy) with the properties that $\Delta S_{gen}(W_1, W_2) = S_{gen}(W_1) - S_{gen}(W_2)$. For $W_1 \subset W_2$, we have $S_{gen}(W_1) < S_{gen}(W_2)$ and for any $W_1,W_2$, we have
    \[
    S_{gen}(W_1) + S_{gen}(W_2) \ge S_{gen}(W_1 \wedge W_2) + S_{gen}(W_1 \vee W_2)~.
    \] 
\end{enumerate}
The last property in particular is highly reminiscent of the strong subadditivity property obeyed by the entropies of quantum subsystems. This similarity motivates the question of whether we can identify $S_{gen}$ with some entropic quantity associated an underlying algebra in such a way that the inequality would follow from an algebraic version of strong subadditivity. 
This is essentially the question of how the Ryu-Takayanagi formula extends to this more general setting. We show that for generalized entanglement wedges with a finite generalized entropy, the assignment\footnote{Here, $S(\omega_W|\mathcal{A}_W)$ can be understood as the entropy of the state $\omega_W$ on the algebra $\mathcal{A}_W$ minus the entropy of the maximum entropy state on this algebra, $\Omega$ is some arbitrary larger algebra that includes $\mathcal{A}_W$ with finite index, ${\rm Ind}(E_{\Omega \to \mathcal{A}_W})$, which is a measure of the size of the algebra $\Omega$ relative to $\mathcal{A}_W$, and $K_\Omega$ is a constant that depends on the choice of algebra $\Omega$. We will explain all of these things in section \ref{sec:gen_ent_wedge}.}
\begin{equation}
\label{prop}
S_{gen}(W) = S(\omega_W|\mathcal{A}_W) - \log {\rm Ind}(E_{\Omega \to \mathcal{A}_W}) + K_\Omega~,
\end{equation}
allows us to derive both the strong subaddivity property and the monotonicity property of $S_{gen}$ for generalized entanglement wedges from inequalities obeyed by algebraic entropies. Our detailed discussion of properties of BP wedges, the proposed map to algebras/states, and the postulated properties of this map may be found in section \ref{sec:gen_ent_wedge}, with the relevant algebraic background explained in the appendix.  

As we describe in section \ref{sec:BP_AdS}, the proposed map between generalized entanglement wedges and algebras would extend the holographic dictionary even in standard AdS/CFT where the generalized entanglement wedges include certain bounded regions in the bulk as well as regions intersecting the conformal boundary that are more general than usual entanglement wedges. In the AdS/CFT context, the identification of generalized entanglement wedges with subalgebras is very reminiscent of the recent Leutheusser-Liu (LL) subregion-subalgebra duality proposal \cite{Leutheusser:2022bgi}, but the context is somewhat different. LL consider a strict large $N$ limit where the bulk theory becomes free quantum field theory on a fixed gravitational background and all gravitational entropies diverge. Further, LL argue that all causally complete subregions of the bulk should be associated with emergent type $III_1$ von Neumann algebras in the boundary theory. On the other hand, in BP, bulk gravitational effects are included (at least perturbatively). They consider wedges with finite generalized entropies and these wedges are required to obey an additional entropy extremality condition: each generalized entanglement wedge minimizes the generalized entropy among all regions that include it. If these wedges have an associated subalgebra, it should be a type $I$ or type $II$ von Neumann algebra\footnote{For a short refresher on the type classification of von Neumann algebras and their associated properties, we refer to appendix \ref{app:algebraicQI}.} to accommodate the finite entropy.\footnote{A subtlety here is that the generalized entropies are only finite for nonzero $G$ (finite $N$ in the AdS/CFT context), while we only have well-defined subregions in the semiclassical (large $N$) limit. One possibility is that the algebras continue to exist at finite $N$ and have entropies that are related to the generalized entropies according to (\ref{prop}). Alternatively, it may be that the algebras may strictly exist only in a large $N$ limit where the entropies are formally infinite, and that (\ref{prop}) should be interpreted as an equivalence either between regulated quantities or between coefficients of entropies in a perturbative $1/N$ expansion. We thank Elliott Gesteau for discussions on this point.}

Even in the AdS/CFT context, it is difficult to evaluate directly whether the proposed map from BP wedges to subalgebras exists. However, as we discuss in section \ref{sec:RTN}, we can make progress in the more tangible context of random tensor network toy models for AdS/CFT.\footnote{Tensor networks were also discussed as part of the initial motivation for BP wedges in \cite{Bousso:2022hlz}. The random tensor network model was also used in \cite{Kaya2025HollowGrams} along with more general path integral arguments to give evidence for the BP proposal.}
Here, we define a set of tensor network subregions analogous to the generalized entanglement wedges and show that for each of these regions, we have a map from a Hilbert space associated with the region to the boundary Hilbert space that becomes isometric in the limit of large bond dimension. We explain how such a map allows us to assign various possible subalgebras of observables to the region. An interesting feature is that if we want the assigned subalgebra to be a von Neumann factor, there are certain choices that must be made in the construction of this boundary algebra, so that we do not just have a single subalgebra assigned to a region but rather a whole family related by a unitary equivalence. This is likely related to the quantum error correcting features of the holographic map and might suggest that there should be a similar non-uniqueness in the map (\ref{map}) in the full gravity context. 

In section \ref{sec:grav_alg}, we comment on the relation of our work to the various recent works that directly construct gravitational algebras associated with certain spacetime regions. 
We conclude in section \ref{sec:fut_dir} with a discussion of some future directions and relations to other work. The material in section \ref{sec:discussion} will be discussed in significantly more detail in an upcoming work \cite{toappear}.

As we will emphasize below, the proposals presented in this work are rather speculative and should be viewed as preliminary ideas subject to refinement and testing rather than a finished product. We hope that the intriguing connections between BP wedge properties and algebra properties presented here may hint at aspects of a more complete picture and promote further useful discussion. 

\section{Generalized entanglement wedges and algebras}

\label{sec:gen_ent_wedge}

Throughout this paper, we will consider gravitational theories in a semiclassical limit where we have well-defined spacetime regions (fluctuations of geometry are perturbative) but the generalized gravitational entropies are UV-finite. 

In this section, we begin by reviewing the definition and various properties of the generalized entanglement wedges of Bousso and Penington, restricting to the case of spatial subsets of a time-reflection-invariant slice $\Sigma$ in a static or time-reflection-symmetric spacetime.

\paragraph{Definition:} {\it A {\bf generalized entanglement wedge}  (or BP wedge) $W$ is a regular open subset\footnote{A regular open set is an open set that is the interior of its closure.} of $\Sigma$ with the property that no region containing $W$ and having the same conformal boundary as $W$ has a smaller generalized entropy.}

Here, the generalized entropy of a region is defined by the usual combination
\[
S_{gen}(W) = {\rm Area(\partial W) \over 4 G} + S_{QFT}(W) \; ,
\]
of the boundary area and the quantum field theory entropy of the region \cite{Lewkowycz2013, Engelhardt2014}. It is believed that this combination is UV-finite, since the normally divergent QFT entropy is regulated by the finite Planck scale in the gravitational theory \cite{Susskind:1994sm}.\footnote{See \cite{Gesteau:2023hbq} for a recent discussion of this statement from an algebraic perspective.}

We would like to explore the following: 
\begin{framed}
{\bf Hypothesis:} For each generalized entanglement wedge $W$, we can associate some subalgebra of observables ${\cal A}_W$ in the underlying description together with an associated state. That is, we have a map
\begin{equation}
    \label{map}
    W \to ({\cal A}_W, \omega_W) \; .
\end{equation}
\end{framed}
\noindent
We recall that a state in the algebraic sense is a linear, positive map from the algebra to $\mathbb{C}$ with $\omega_W(1) = 1$. Below, we will be more explicit about the type of algebra and the properties of the state.

\subsection{Basic properties}

We will now motivate this hypothesis by recalling various properties of BP wedges and showing that in each case, there is a corresponding property of subalgebras that can be naturally associated.
\paragraph{Inclusion}
The set of generalized entanglement wedges has a natural partial order based on inclusion. This is also true for subalgebras. Thus, it is natural to propose that the map (\ref{map}) obeys
\begin{framed}
\begin{enumerate}
\item 
$W_1 \subset W_2 \iff \mathcal{A}_{W_1} \subset  \mathcal{A}_{W_2}~. $
\end{enumerate}
\end{framed}
\paragraph{Spacelike separation / commutation}
Certain pairs of BP wedges are spacelike separated. Similarly, certain pairs of subalgebras commute with each other. For spatial subsystems of a quantum field theory, subalgebras commute if and only if the corresponding regions are spacelike separated. Thus, it is plausible that the map (\ref{map}) also satisfies
\begin{framed}
\begin{enumerate}
\setcounter{enumi}{1}
\item 
$W_1, W_2 {\rm \; spacelike \; separated}  \iff \mathcal{A}_{W_1} \subset  \mathcal{A}_{W_2}'~,$
\end{enumerate}
\end{framed}
\noindent
where the inclusion relation holds in any larger algebra that contains both $\mathcal{A}_{W_1}$ and $\mathcal{A}_{W_2}$ and the commutant algebra $\mathcal{A}_{W_2}'$ is defined within this larger algebra.

\paragraph{Intersections}

Given any two BP wedges $W_1$ and $W_2$, we can consider their intersection $W_1 \cap W_2$. It turns out that 
\begin{itemize}
    \item {\it The intersection of any two generalized entanglement wedges is a generalized entanglement wedge.}
\end{itemize}
Since this was not shown previously as far as we are aware, we include a proof here. We begin with the following Lemma:
\begin{itemize}
    \item {\it 
If $W$ is a generalized entanglement wedge and $a$ is regular open subset of $\Sigma$, we have
\begin{equation}
\label{BPmon}
 S_{gen}(a \cap W) \le S_{gen}(a)~.
\end{equation}}
\end{itemize}
To see this latter property, we first note that for any regular open sets $a$ and $b$, 
\[
S_{gen}(a) + S_{gen}(b) \ge S_{gen}(a \cup b) + S_{gen}(a \cap b)~.
\]
This follows separately for the area term, by the submodularity property of boundary areas,\footnote{For any infinitesimal part of the boundary of $a$, we can either have that neither side is in $b$, only the $a$ side is in $b$, or both sides are in $b$, or only the non-$a$ side is in $b$. In the first three cases, this area contributes in the same way on both sides of the inequality, while in the last case, this area only contributes on the left side. A similar argument applies to any part of the boundary of $b$.} and for the bulk entanglement term, by strong subadditivity.

Taking $b=W$ to be a generalized entanglement wedge, we have
\[
S_{gen}(W \cup a) \ge S_{gen}(W)~,
\]
by the defining property of generalized entanglement wedges, so 
\[
S_{gen}(a) + S_{gen}(W) \ge S_{gen}(a \cup W) + S_{gen}(a \cap W) \ge S_{gen}(W) + S_{gen}(a \cap W)  \; .
\]
Now using (\ref{BPmon}), we can say that if $A$ and $B$ are generalized entanglement wedges and $C$ is any set that contains their intersection,
\[
S_{gen}(A \cap B) = S_{gen}(C \cap A \cap B) = S_{gen}((C \cap A) \cap B) \le S_{gen}(C \cap A) \le S_{gen}(C)~.
\]
This establishes that $A \cap B$ is also a generalized entanglement wedge.

On the algebra side, it is a well-known property that the intersection of two algebras is an algebra (this also applies to von Neumann algebras). Thus, it seems natural to assert that the map (\ref{map}) should satisfy
\begin{framed}
\begin{enumerate}
\setcounter{enumi}{2}
\item
    $\mathcal{A}_{W_1 \cap W_2} = \mathcal{A}_{W_1} \wedge \mathcal{A}_{W_2} \; .$
\end{enumerate}
\end{framed}
A weaker version of this that must hold assuming property 1 is that 
\begin{enumerate}
\setcounter{enumi}{2}
\item (weaker version)
    $\mathcal{A}_{W_1 \cap W_2} \subset \mathcal{A}_{W_1} \wedge \mathcal{A}_{W_2} \; .$
\end{enumerate}
\noindent
This follows since $W_1 \cap W_2$ is in both $W_1$ and $W_2$ so that according to property 1, the corresponding algebra is in both $\mathcal{A}_{W_1}$ and $\mathcal{A}_{W_2}$. 

\paragraph{Unions}

The union of BP wedges is generally not a BP wedge, but there is a canonical way to assign a BP wedge to any union of BP wedges by taking the intersection of all BP wedges that include $\cup_i W_i$. We refer to this as the {\it join} of the wedges and denote it as $\vee_i W_i$. Equivalently, $\vee_i W_i$ is BP wedge that includes $\cup_i W_i$ and has the smallest generalized entropy.

Parallel comments apply to subalgebras. The union of subalgebras is generally not a subalgebra, but there is a canonical way to assign a subalgebra to any union of subalgebras by taking the intersection of all subalgebras that include $\cup_i \mathcal{A}_i$, or equivalently, the algebra generated by the operators in $\cup_i \mathcal{A}_i$. This is also called the join of the subalgebras and denoted $\vee_i \mathcal{A}_i$.

Based on these similarities, a natural proposal is that that the map (\ref{map}) has the property  
\begin{framed}
\begin{enumerate}
\setcounter{enumi}{3}
\item $\mathcal{A}_{W_1 \vee W_2} = \mathcal{A}_{W_1} \vee \mathcal{A}_{W_2}~.$
\end{enumerate}
\end{framed}
In order to reproduce the generalized entropy inequalities algebraically, it turns out that we will only require a weaker version of this proposal, namely
\begin{framed}
\begin{enumerate}
\setcounter{enumi}{3}
\item  (weaker version) $\mathcal{A}_{W_1 \vee W_2} \supset \mathcal{A}_{W_1} \vee \mathcal{A}_{W_2}~.$
\end{enumerate}
\end{framed}
In section \ref{sec:grav_alg} below, we recall discussions of the direct construction of gravitational algebras associated with spacetime regions. In these constructions, it is not clear that the stronger version of property 4 holds, so we also include the weaker alternative here as a possibility. 

It is worthwhile to note that while properties 1 and 2 hold for algebras of local observables in QFT, properties 3 and 4 (in their strong form) may be violated for theories containing superselection sectors (see section 2 of \cite{Casini:2019kex} and references there in). Holographic theories are examples of theories with superselection sectors and hence, at least for generalized wedges in asymptotically AdS spacetimes, the algebras hypothesized here might include non-local operators in order to satisfy properties 1-4. We will see an illustration of this feature in the tensor network model discussed in \cref{sec:discussion}.

\subsection{Generalized entropy and a generalized RT formula}

We now turn to the quantitative properties of the generalized entanglement wedges, seeking an algebraic interpretation for the generalized entropy. We begin with the case of bounded regions, discussing more general regions in section \ref{sec:conformal_bdy}.

For a bounded BP wedge $W$, we can assign a finite generalized entropy $S_{gen}(W)$ with the properties \cite{Bousso:2022hlz}:
\begin{itemize}
    \item 
     {\bf Monotonicity}: $W_1 \subset W_2 \implies S_{gen}(W_1) < S_{gen}(W_2)~.$
    \item
     {\bf Strong subadditivity}: For any $W_1,W_2$, $S_{gen}(W_1) + S_{gen}(W_2) \ge S_{gen}(W_1 \cap W_2) + S_{gen}(W_1 \vee W_2)~.$
\end{itemize}
The latter property follows from Theorem 3.17 in \cite{Bousso:2022hlz} by choosing $a = W_1 - W_1 \cap W_2$, $b = W_1 \cap W_2$, $c = W_2 - W_1 \cap W_2$, and noting that $W_1 \vee W_2 = E(W_1 \cup W_2)$ in the notation of \cite{Bousso:2022hlz}. 

For ordinary entanglement wedges, the strong subadditivity property follows from strong subadditivity property for quantum subsystems in the (regulated) underlying field theory, together with the quantum RT formula
\begin{equation}
\label{RT}
S_{gen}(EW(A)) = S_{vN}(A)~, \; 
\end{equation}
where the right hand side is the von Neumann entropy of the boundary subsystem $A$ associated with entanglement wedge $EW(A)$ \cite{Headrick:2007km}.

We would like to understand whether there is a generalization of the RT-formula (\ref{RT}) that identifies the generalized entropy of a generalized entanglement wedge with some entropic quantity associated with the underlying subalgebra such that the monotonicity and strong subadditivity properties above follow from properties of the algebraic entropy.\footnote{In the end, the formula we obtain will generalize a slightly different version of (\ref{RT}) where the generalized entropy of an ordinary entanglement wedge includes the (divergent) area of the conformal boundary of $A$ and the right hand side includes a divergent coefficient times the volume of $A$. A regulated version of this entropy obeys the monotonicity property, while the regulated von Neumann entropy does not.}

\subsection{Algebraic entropies}

To proceed, we will need to recall how entropy generalizes from the usual subsystem case (von Neumann entropy of the reduced density operator associated with a tensor factor) to more general subalgebras. 

For a general von Neumann algebra $\mathcal{A}$, we can always define the relative entropy $S(\omega|| \sigma)$ between two states $\omega$ and $\sigma$ via Tomita-Takesaki theory\footnote{We refer the interested reader to \cite{Takesaki:1970aki, TakesakiII} for more background on the Tomita-Takesaki modular theory.} \cite{Araki:1976zv, Araki:1977zsq}. In the type $I_n$ or type $II_1$ settings, we have a canonical finite trace $\tau$ with $\tau(1) = 1$. The trace is a map from the algebra of operators to $\mathbb{C}$, satisfying the properties that define the algebraic version of a quantum state. This tracial state provides a canonical reference state to which other states can be compared.\footnote{We will usually assume this trace to be \textit{faithful}, meaning that it is strictly positive for every non-zero positive operator, and \textit{normal}, meaning that it is continuous on the space of operators in some appropriate sense (the normality condition being non-trivial only in the infinite-dimensional setting).} In this case, we can define an entropy \cite{Segal1960ANO, Longo:2022lod}
\[
S(\omega) = - S(\omega||\tau)~,
\]
as minus the relative entropy between the state $\omega$ and the reference state $\tau$. By the positivity of relative entropy, this is always negative, so it has the interpretation of the difference in entropy between the state $\omega$ and the maximum entropy state $\tau$ (equal to the maximally mixed state in the standard case).

We can write more explicit formulae for entropy and relative entropy in this case with a finite trace by defining the density operator $\rho_\omega$ associated with a state $\omega$ as the unique\footnote{The uniqueness of the density operator $\rho_\omega$ is a consequence of faithfulness of the trace.} operator satisfying
\[
\omega(a) = \tau(\rho_\omega a)~, \qquad \forall a \in \mathcal{A} \; .
\]
In terms of this, the relative entropy between states $\omega$ and $\sigma$ is 
\[
S(\omega || \sigma) = \tau(\rho_\omega \log \rho_\omega - \rho_\omega \log \rho_\sigma)~,
\]
and using that $\rho_{\tau}=1$, the entropy of a state $\omega$ is
\[
S(\omega) = -\tau(\rho_\omega \log \rho_\omega) \; .
\]
Because the trace here is normalized as $\tau(1)=1$, this differs from the usual subsystem von Neumann entropy in the standard case where $\mathcal{A}$ is the type $I_D$ algebra of operators acting on a $D$-dimensional Hilbert space, giving:
\[
S(\omega) = S_{vN}(\rho_{\omega}) - \log D \; .
\]

\paragraph{Coarse-graining via conditional expectations}
We will also need an algebraic notion of coarse-graining a state to a subalgebra $\mathcal{B} \subset \mathcal{A}$. When $\mathcal{A}$ has a trace $\tau$, we can use it to define an inner product $\langle a,b \rangle = \tau(a^* b)$, and then use this inner product to orthogonally project any operator in $\mathcal{A}$ to an operator in $\mathcal{B}$. This projection defines a trace-preserving {\it conditional expectation} $E:\mathcal{A} \to \mathcal{B}$, an object that we review and define more generally in section \ref{sec:cond_expectation} of the appendix.

The conditional expectation has the property that for any $a \in \mathcal{A}$, $b \in \mathcal{B}$, 
\[
\tau(E(a) b) = \tau(a b) \; .
\]
In particular, applied to the density operator $\rho_\mathcal{A}$ associated with a state $\omega_\mathcal{A}$, we have 
\[
\tau(E(\rho_\mathcal{A}) b) = \tau(\rho_\mathcal{A} b) \; .
\]
so $\rho_\mathcal{B} \equiv E(\rho_\mathcal{A})$ has the properties usually associated with a reduced density operator. In fact $\rho_\mathcal{B}$ (or more precisely its map back to the original algebra by inclusion) is the density operator associated with the state $\omega \circ E: \mathcal{A} \to \mathbb{C}$  which can be interpreted as the original state $\omega$ coarse-grained to preserve all the information about observables on the subalgebra but discarding as much other information as possible. 

\paragraph{Entropy increases under coarse-graining}

For a state $\omega_\mathcal{A}$ on $\mathcal{A}$, we will indicate the entropy of the coarse-grained state $\omega \circ E_{\mathcal{A} \to \mathcal{B}}$ by
\[
S(\omega | \mathcal{B}) = S(\omega \circ E_{\mathcal{A} \to \mathcal{B}}) =  - \tau(\rho_\mathcal{B} \log \rho_\mathcal{B}) \; .
\]
Intuitively, we expect that entropy should increase under coarse-graining. This is correct, since the entropy difference between the coarse-grained state and the original state can be expressed as a relative entropy:
\begin{equation}
\label{badineq}
S(\omega | \mathcal{B}) - S(\omega | \mathcal{A}) = S(\omega || \omega \circ E) \ge 0 \; .
\end{equation}

\subsection{An algebraic RT formula}
\label{sec:alg_RT}

We now return to the question of whether there is some entropic quantity associated with the algebra $\mathcal{A}_W$ that should be identified with the generalized entropy of a BP wedge $W$. 

The standard quantum RT formula posits an identification
\begin{equation}
\label{RT2}
S_{gen}(EW(A)) = S_{vN}(A)~,
\end{equation}
between generalized entropy of an ordinary entanglement wedge $EW(A)$ and the von Neumann entropy of the associated boundary subsystem $A$, though the right side is often divergent without regularization.

In the finite-dimensional case when we consider a matrix algebra of dimension $D_A^2$, the von Neumann entropy and the algebraic entropy are related by addition of a state-independent constant
\[
S_{vN}(A) = S(\omega|\mathcal{A}_A) + \log D_A \; ,
\]
so more generally, we might expect an identification of the form
\[
S_{gen}(W) = S(\omega|\mathcal{A}_W) + \dots
\]
for some state-independent quantity that replaces subsystem dimension. 

To proceed, we recall that the generalized entropy satisfies the monotonicity property
\[
S_{gen}(W_B) - S_{gen}(W_A) \le 0~, \qquad W_B \subset W_A \; .
\]
This looks a little like the coarse-graining relation (\ref{badineq}),
\[
S(\omega | \mathcal{B}) - S(\omega | \mathcal{A}) \ge 0~, \qquad \mathcal{B} \subset \mathcal{A}~,
\]
but here the monotonicity has the opposite behavior: the generalized entropy is larger for the larger subsystem, while the algebraic entropy is larger for the smaller subsystem. 

It turns out that there is an inequality involving the algebraic entropies where the inequality goes in the right direction \cite{Longo:2022lod, longo1989index}\footnote{An application of this formula also appears in \cite{Casini:2019kex} with regards to the calculation of relative entropies in QFT with superselection sectors.} :
\begin{equation}
\label{coarsebound}
S(\omega | \mathcal{B}) - S(\omega | \mathcal{A}) \le \log {\rm Ind}(E_{\mathcal{A} \to \mathcal{B}}) \; .
\end{equation}
This expresses a state-independent upper bound for the increase in entropy under coarse-graining. On the right hand side, the {\it index} ${\rm Ind}(E_{\mathcal{A} \to \mathcal{B}})$ of the conditional expectation $E$ is a measure of the size of the algebra $\mathcal{A}$ relative to the subalgebra $\mathcal{B}$.\footnote{See section \ref{sec:index} for a full discussion of the index.
Even in the type $II_1$ context where conditional expectations exist for inclusions $\mathcal{A} \subset \mathcal{B}$, the index need not be finite. Below (property 9) we will add an explicit assumption that region inclusions where we have a finite difference of generalized entropy, the algebra inclusions have conditional expectations with finite index. This is a stringent requirement, and we view our construction as a first proposal that achieves the desired behavior. It may well be that, in a more realistic description, this structure will need to be refined; see, for example, the recent discussion of generalized conditional expectations in \cite{AliAhmad:2024saq}.} 
For example, if  $\mathcal{A}$ and $\mathcal{B}$ are the algebras of operators acting on a $D_A$-dimensional Hilbert space and an $D_B$-dimensional tensor factor subsystem, we have simply
\[
{\rm Ind}(E_{\mathcal{A} \to \mathcal{B}}) = {\dim{\mathcal{A}} \over \dim{\mathcal{B}}} = {D_A^2 \over D_B^2} \; .
\]

In the case where $\mathcal{A} \subset \mathcal{B} \subset \Omega$ and each of these algebras is a von Neumann factor\footnote{A von Neumann factor is a von Neumann subalgebra with a trivial center, that is, where only multiples of the identity operator commute with all operators in the algebra.} we have the relation\footnote{In general when we are not assuming factors, the left side is always smaller than or equal to the right side.}
\[
{\rm Ind}(E_{\Omega \to \mathcal{B}}) = {\rm Ind}(E_{\Omega \to \mathcal{A}}) {\rm Ind}(E_{\mathcal{A} \to \mathcal{B}}) \; ,
\]
so the inequality (\ref{coarsebound}) can be rearranged to give
\[
[S(\omega | \mathcal{A}) - \log {\rm Ind}(E_{\Omega \to \mathcal{A}})] - [S(\omega | \mathcal{B}) - \log {\rm Ind}(E_{\Omega \to \mathcal{B}})] \ge 0  \; .
\]
Thus, the following identifications would provide an algebraic origin for the BP monotonicity property:
\begin{framed}
\begin{enumerate}
\setcounter{enumi}{4}
\item For a generalized entanglement wedge $W$, the algebra $\mathcal{A}_W$ is a von Neumann factor.
\item 
    For a generalized entanglement wedge $W$ with finite generalized entropy, we have  
\begin{equation}
\label{AlgRT}
S_{gen}(W) = S(\omega | \mathcal{A}_W) - \log {\rm Ind}(E_{\Omega \to \mathcal{A}_W}) + K_\Omega~,
\end{equation}
where $\Omega$ is an arbitrary larger algebra in which $\mathcal{A}_W$ sits with finite index, and $K_\Omega$ is a constant that depends only on the choice of $\Omega$.
\end{enumerate}
\end{framed}
These identifications lead to a simple identification for differences of generalized entropies. For BP wedges $W_B \subset W_A$ with associated algebras $\mathcal{B}$ and $\mathcal{A}$, state $\omega$ on $\mathcal{A}$ and conditional expectation $E = E_{\mathcal{A} \to \mathcal{B}}$, we have
\begin{equation}
\label{eq:Sdiff}
S_{gen}(W_A) - S_{gen}(W_B) =S(\omega | \mathcal{A}) - S(\omega |\mathcal{B}) + \log {\rm Ind} 
\,E = -S(\omega || \omega \circ E) + \log {\rm Ind} \,E \; .
\end{equation}
The last expression will be useful below since it can apply even when individual entropies diverge.

To understand better the proposed identification (\ref{AlgRT}), consider the right side in the case where $\mathcal{A}_W$ is the algebra of operators on a tensor factor Hilbert space ${\cal H}_A$ of dimension $D_A$ and $\Omega$ is the algebra of operators on some larger Hilbert space ${\cal H}_\Omega$ of dimension $D_\Omega$ that includes the factor ${\cal H}_A$. We might have in mind that ${\cal H}_A$ is associated with a regulated quantum field theory on a subsystem $A$ and ${\cal H}_\Omega$ is associated with the quantum field theory on a larger subsystem that contains $A$. 
Then 
\[
S(\omega | \mathcal{A}_W) = S_{vN}(\rho_A) - \log D_A
\]
and 
\[
{\rm Ind}(E_{\Omega \to \mathcal{A}_W}) = {D_\Omega^2 \over D_A^2} \; .
\]
We can eliminate the dependence on our choice of ${\cal H}_\Omega$ by taking $K_{\Omega}=\log D_{\Omega}^2$, and this gives
\[
S(\omega | \mathcal{A}_W) - \log {\rm Ind}(E_{\Omega \to \mathcal{A}_W}) + K_\Omega = S_{vN}(\rho_A) + \log D_A \; .
\]
This does not reduce to the right hand side of the RT formula (\ref{RT2}), so it may seem that our proposed identification does not actually generalize the RT formula. However, the extra term $\log D_A$ has a natural interpretation. If in the RT formula (\ref{RT2}) we regularize by replacing the region $EW(A)$ with a bounded region $EW(A)_\epsilon$ of the bulk geometry that is the part of $EW(A)$ inside some cutoff surface $\Sigma_\epsilon$, then the boundary of $EW(A)_\epsilon$ contains a part of $\Sigma_\epsilon$, and the generalized entropy of $EW(A)_\epsilon$ includes a term corresponding to the area of this surface. The term $\log D_A$ in our proposed formula accounts for this additional area. The extra contribution to $S_{gen}$ and the corresponding extra term in the entropy are necessary so that $S_{gen}(EW(A)_\epsilon)$ behaves continuously in the limit where $\epsilon \to 0$. The extra divergent term that we are adding is state-independent and cancels out in any combinations of entropies that yield finite results. 

\paragraph{Strong subadditivity}

In addition to monotonicity, the generalized entropy satisfies the strong subadditivity property. We would like to understand whether this follows from our identification (\ref{AlgRT}) and some algebraic version of strong subadditivity. 

The first question is how the standard subsystem strong subadditivity result generalizes to subalgebras. We review a number of relevant results in Appendix \ref{sec:alg_SSA}. One sufficiently general result is \cite{petz1991certain}:
\paragraph{Theorem}{\it Suppose $\mathcal{A}$ is a von Neumann algebra with a faithful normal finite trace $\tau$  and subalgebras $\mathcal{B}$ and $\mathcal{C}$ with trace-preserving conditional expectations $E_{\mathcal{A} \to \mathcal{B}}$ and $E_{\mathcal{A} \to \mathcal{C}}$ that commute. Then for any state $\omega$
\[
S(\omega|\mathcal{B}) + S(\omega| \mathcal{C}) \ge  S(\omega | \mathcal{A}) + S(\omega | \mathcal{B} \wedge \mathcal{C})   \; .
\]}
The commuting of the conditional expectations is known as the \emph{commuting square condition}. The analogous statement for the inclusion of commutant algebras is called the \emph{co-commuting square condition}. In this case, we say that the commuting square is \emph{non-degenerate}. This also implies a useful relation between the indices \cite{sano1996commuting}:\footnote{Here we are assuming the various algebras are factors, though the relation holds more generally with a center-valued index that can be defined in the non-factor case.} 
\[
{\rm Ind}(E_{\mathcal{A} \to \mathcal{C}}) = {\rm Ind}(E_{\mathcal{B} \to \mathcal{B} \wedge \mathcal{C}}) \; .
\]
We refer to Appendix \ref{sec:index} for more details. Combining this with the algebraic strong subadditivity relation above gives (assuming the conditions of the theorem and that the various algebras are factors)
\[
S(\omega|\mathcal{B}) - S(\omega | \mathcal{B} \wedge \mathcal{C}) + \log {\rm Ind}(E_{\mathcal{B} \to \mathcal{B} \wedge \mathcal{C}}) \ge  S(\omega | \mathcal{A}) -  S(\omega| \mathcal{C})  + \log {\rm Ind}(E_{\mathcal{A} \to \mathcal{C}})  \; .
\]
Now consider generalized entanglement wedges $W_1$ and $W_2$ and take $\mathcal{B} = \mathcal{A}_{W_1}$, $\mathcal{C} = \mathcal{A}_{W_2}$, and $\mathcal{A} = \mathcal{A}_{W_1 \vee W_2}$. Again assuming the conditions of the theorem hold for these algebras, the previous inequality gives 
\[
S(\omega|\mathcal{A}_{W_1}) - S(\omega | \mathcal{A}_{W_1} \wedge \mathcal{A}_{W_2}) + \log {\rm Ind}(E_{\mathcal{A}_{W_1} \to \mathcal{A}_{W_1} \wedge \mathcal{A}_{W_2}}) \ge  S(\omega | \mathcal{A}_{W_1 \vee W_2}) -  S(\omega| \mathcal{A}_{W_2})  + \log {\rm Ind}(E_{\mathcal{A}_{W_1 \vee W_2} \to \mathcal{A}_{W_2}})  \; .
\]
Using property 3, we can replace $\mathcal{A}_{W_1} \wedge \mathcal{A}_{W_2}$ with $\mathcal{A}_{W_1 \wedge W_2}$ to get\footnote{For this step, we actually only require the weaker result that $\mathcal{A}_{W_1} \wedge \mathcal{A}_{W_2}$ is a factor that includes $\mathcal{A}_{W_1 \wedge W_2}$ with a finite index conditional expectation. In this case, we have $0 \le S(\omega|\mathcal{A}_{W_1} \wedge \mathcal{A}_{W_2}) - S(\omega|\mathcal{A}_{W_1 \wedge W_2}) + \log {\rm Ind} E_{\mathcal{A}_{W_1} \wedge \mathcal{A}_{W_2} \to \mathcal{A}_{W_1 \wedge W_2}} \ge 0$ and $\log {\rm Ind} E_{\mathcal{A}_{W_1} \wedge \mathcal{A}_{W_2} \to \mathcal{A}_{W_1 \wedge W_2}} =  \log {\rm Ind} E_{\mathcal{A}_{W_1} \to \mathcal{A}_{W_1 \wedge W_2}} - \log {\rm Ind} E_{\mathcal{A}_{W_1} \to \mathcal{A}_{W_1} \wedge \mathcal{A}_{W_2}}$ so that $- S(\omega|\mathcal{A}_{W_1 \wedge W_2})      + \log {\rm Ind} E_{\mathcal{A}_{W_1} \to \mathcal{A}_{W_1 \wedge W_2}} \ge - S(\omega|\mathcal{A}_{W_1} \wedge \mathcal{A}_{W_2}) + \log {\rm Ind} E_{\mathcal{A}_{W_1} \to \mathcal{A}_{W_1} \wedge \mathcal{A}_{W_2}}$ that allows us to complete this step in the derivation.}
\[
S(\omega|\mathcal{A}_{W_1}) - S(\omega | \mathcal{A}_{W_1 \wedge W_2}) + \log {\rm Ind}(E_{\mathcal{A}_{W_1} \to \mathcal{A}_{W_1 \wedge W_2}}) \ge  S(\omega | \mathcal{A}_{W_1 \vee W_2}) -  S(\omega| \mathcal{A}_{W_2})  + \log {\rm Ind}(E_{\mathcal{A}_{W_1 \vee W_2} \to \mathcal{A}_{W_2}})  \; .
\]
Finally, the identification \cref{eq:Sdiff} gives 
\[
S_{gen}(W_1) - S_{gen}(W_1 \cap W_2) \ge   S_{gen}(W_1 \vee W_2) - S_{gen}(W_2)~,
\]
which is equivalent to the desired strong subadditivity relation.

Since we needed the commuting and co-commuting square conditions, we should add this on to our list of postulated properties for the map (\ref{map}):
\begin{framed}
\begin{enumerate}
\setcounter{enumi}{6}
\item For generalized entanglement wedges $W_1$ and $W_2$, there exist commuting conditional expectations $E_1: \mathcal{A}_{W_1 \vee W_2} \to \mathcal{A}_{W_1}$ and $E_2: \mathcal{A}_{W_1 \vee W_2} \to \mathcal{A}_{W_2}$ such that the co-commuting square condition is satisfied.
\end{enumerate}
\end{framed}

\subsection{Entanglement wedges with a conformal boundary}

\label{sec:conformal_bdy}

For generalized entanglement wedges that intersect the conformal boundary, the generalized entropy diverges, so our entropy formula above is not useful. But in some cases, two entanglement wedges with the same conformal boundary may have a finite difference of generalized entropy. Even with the same conformal boundary, the entropy difference can still diverge if $\partial W_1$ and $\partial W_2$ approach this boundary in different ways.\footnote{For example, in AdS${}_3$, we can consider wedges that intersect the same boundary interval but at different angles.} We will refer to wedges that have the same conformal boundary but also finite entropy differences as being in the same class. These classes are closed under intersections and joins, and the intersection of all wedges in a class is the ordinary entanglement wedge associated with the conformal boundary of any one of the wedges. This suggests:
\begin{framed}
    \begin{enumerate}
    \setcounter{enumi}{7}
        \item 
        For a given class of BP-wedges characterized by the property of finite entropy differences, the corresponding algebras form a sublattice, a set closed under (finite) algebra intersections and joins. The intersection of all algebras in the sublattice is the algebra associated with the usual entanglement wedge associated with the conformal boundary of one of the wedges. This may or may not be in the sublattice.
    \end{enumerate}
\end{framed}
When the generalized entropies of regions $B \subset A$ are finite, the entropy difference can be expressed via (\ref{eq:Sdiff}) in terms of a relative entropy and an index, both constructed from the conditional expectation $E: \mathcal{A} \to \mathcal{B}$. The resulting expression on the right side of (\ref{eq:Sdiff}) makes sense and is finite more generally whenever we have a finite-index conditional expectation between the two algebras. Thus, it seems natural to identify the property of having finite generalized entropy difference with the property of there existing a finite-index conditional expectation between the corresponding algebras. This leads to the following proposal:
\begin{framed}
    \begin{enumerate}
    \setcounter{enumi}{8}
    \item
        BP-wedges $W_A$ and $W_B$ sharing the same conformal boundary and having a finite difference of generalized entropies are associated with algebras $\mathcal{A}$ and $\mathcal{B}$ for which there are finite index conditional expectations $E_{\mathcal{C} \to \mathcal{A}}$ and $E_{\mathcal{C} \to \mathcal{B}}$ from the joint algebra $\mathcal{C} = \mathcal{A} \vee \mathcal{B}$. If $\omega$ is the state on $\mathcal{C}$, we have
\begin{equation}
\label{Sdiff2}
S_{gen}(W_A) - S_{gen}(W_B) = [S(\omega|| \omega \circ E_{\mathcal{C} \to \mathcal{A}}) - \log {\rm Ind} E_{\mathcal{C} \to \mathcal{A}}] - [S(\omega|| \omega \circ E_{\mathcal{C} \to \mathcal{B}}) - \log {\rm Ind} E_{\mathcal{C} \to \mathcal{B}}] \; .
\end{equation}
    \end{enumerate}
\end{framed}
For $W_B \subset W_A$, the last formula reduces to (\ref{eq:Sdiff}).

\section{Discussion}

\label{sec:discussion}

We have proposed a set of features of a possible map from generalized entanglement wedges to algebras that allows us to give all of the key features of the wedges an algebraic origin. It is entirely possible that some of the suggestions above are too naive. For example, as we describe below based on a tensor network toy model, it may be that a BP wedge is associated with some family of algebras related by unitary equivalence rather than to a single algebra. Thus, we suggest that these proposals should be viewed as a starting point for further investigations. Below, we discuss various ways in which these can be tested.

\subsection{Bousso-Penington wedges in asymptotically AdS spacetimes}

\label{sec:BP_AdS}

For general spacetimes, we do not have any direct way to evaluate whether the proposals of the previous section are true. However, even for asymptotically AdS spacetimes where we have a microscopic description via holography, these proposals extend significantly the established dictionary. Thus it is worthwhile in the AdS/CFT context to evaluate more directly whether generalized entanglement wedges have dual algebras and if so, whether these algebras satisfy the properties that we have shown lead to the Bousso-Penington results. 

For usual entanglement wedges in asymptotically AdS spacetimes, we do have naturally associated algebras, describing the dual field theory observables living in the spacetime region associated with the conformal boundary of the entanglement wedge. For more general BP wedges, we expect that the associated algebras would be of a more general type that would not be associated with a spatial subregion. 

In AdS/CFT, we have a global type I algebra of observables at finite $N$. 
However, the proper context for our discussion in the previous section is likely some large $N$ limit where we consider some particular family of states $|\Psi_N \rangle$ that gives rise to a limiting state with a good semiclassical description (e.g. states created by a Euclidean path integral with a geometry and sources that are fixed as we take $N$ large). The global algebra that is relevant here is the set of operators that remain well-defined in this large $N$ limit. 

These large $N$ algebras of observables have been discussed in detail by Leutheusser and Liu (LL) in \cite{Leutheusser:2021frk, Leutheusser:2022bgi}. In \cite{Leutheusser:2022bgi} these authors propose a subregion-subalgebra duality that is very similar to the one we have discussed in the previous section. However, LL consider a strict large $N$ limit where the resulting bulk physics is free quantum field theory on a fixed gravitational background. Here, it is expected that (as for quantum field theory in general), any causal diamond region (the domain of dependence of some open subset of a Cauchy slice) should have an associated type $III_1$ algebra of observables, and LL propose that all of these type $III_1$ algebras should arise from the large $N$ limit of the boundary algebra. States of type $III$ algebras do not have well-defined finite entropies, and this corresponds to the fact that the strict large $N$ limit corresponds to a $G \to 0$ limit in the bulk where the geometrical contribution $A/(4G)$ to generalized entropy diverges even for finite $A$.

In the Bousso-Penington discussion, there are a couple of key differences. First, the generalized entanglement wedges are not associated with arbitrary causal diamonds, but only regions that obey an extremality condition (that there is no larger wedge with a smaller generalized entropy). Second, the numerical values of generalized entropies play an important role in the BP discussion, so we must be working in a context where these entropies are finite. On the other hand, BP also assume that we are in a context where we can discuss well-defined regions in a gravitational theory. Together, these observations suggest that we are still in a large $N$ limit, but one in which we take into account bulk gravitational effects perturbatively. 

One possible picture is that for a generalized entanglement wedge, the associated large $N$ LL algebra survives $1/N$ corrections to become a type $II$ algebra with finite entropy (for a finite area region), while more general regions not satisfying the extremality condition do not have associated algebras. We will provide some evidence for this picture in a tensor network toy model discussed below. 

In an upcoming work, we will discuss in more detail the generalized entanglement wedges in AdS spacetimes. The case of AdS${}_3$ is particularly simple since we can characterize the generalized entanglement wedges living on a spatial slice very simply:
\begin{itemize}
\item 
{\it A regular open set in $H^2$ is a generalized entanglement wedge if and only if it is a union of geodesically convex open sets with the property that the convex hull of any collection of these sets has a perimeter larger than or equal to than that of the collection of sets.}
\item {\it A geodesically convex connected open set in $H^2$ is equal to the intersection of all the ordinary entanglement wedges that contain it.}
\end{itemize}
Together, these two properties show that any BP wedge in $H^2$ can be defined as a join of intersections of ordinary entanglement wedges, so according to the strong versions of proposals 3 and 4 in section \ref{sec:gen_ent_wedge}, the corresponding algebra would be the join of intersections of algebras associated to ordinary entanglement wedges.\footnote{For more general spacetimes (even AdS${}_4$) it is not true that a connected generalized entanglement wedge can be described as the union of intersection of usual entanglement wedges.} In this case, it seems that the key question is the precise way that the algebra associated with a usual entanglement wedge should be defined in the ``soft'' large $N$ limit where we still include gravitational effects. 

\subsection{Random tensor network toy model}

\label{sec:RTN}

A very tangible setting in which to explore the proposals of section \ref{sec:gen_ent_wedge} is in the context of the random tensor network toy model of AdS/CFT \cite{Hayden:2016cfa}. Here, the geometry of a tensor network takes the place of the bulk gravitational geometry, and the tensor network state associated with the open boundary legs of the network models the state of the holographic field theory that encodes this geometry.\footnote{In \cite{toappear} we will discuss more general tensor networks with bulk states.}

In the tensor network context, the analog of a bulk subregion is a collection of vertices in the network. Removing the vertices in a subregion $A$ opens a set of bulk legs associated with Hilbert space $\mathcal{H}_A$, and the remaining tensor network defines a map $T_A$ from this Hilbert space to the Hilbert space $\mathcal{H}_\partial$ associated to the boundary legs.

In the limit of large bond dimension (which models the large $N$ limit in AdS/CFT), the map $T_A$ becomes an isometry if and only if the region $A$ obeys an extremality condition similar to the one defining BP wedges: we require that there is no larger tensor network subsystem containing $A$ with a smaller ``area'', where area means the number of tensor network legs revealed by removing the vertices in the region.

Having an isometry $T_A: \mathcal{H}_A \to \mathcal{H}_\partial$ allows us to associate to $A$ a subalgebra of the algebra of operators on $\mathcal{H}_\partial$. The simplest construction, based on the map 
\[
\mathcal{O}_A \to T_A \mathcal{O}_A T_A^\dagger~, 
\]
defines a $C^*$-subalgebra $\mathcal{C}_A$ of the boundary algebra. This is a {\it corner} of the algebra $B(\mathcal{H}_\partial)$ of operators on $\mathcal{H}_\partial$ since it can be equivalently defined as the set of operators $P_A B(\mathcal{H}_\partial) P_A$, where $P_A = T_A T_A^\dagger$ is a projector. 

This subalgebra $\mathcal{C}_A$ is not a von Neumann subalgebra of the larger algebra since it does not contain the identity operator in $B(\mathcal{H}_\partial)$. We can generate a von Neumann subalgebra by taking the double-commutant $\mathcal{M}_A \equiv \mathcal{C}_A''$. This is the subalgebra of operators of the form
\[
P {\cal O} P + \lambda (1-P) \; .
\]
for arbitrary $\lambda \in \mathbb{C}$. The algebra $\mathcal{M}_A$ is a von Neumann subalgebra but it is not generally a factor, since operators of the form $a P + b (1-P)$ give a non-trivial commutant. 

We can also define a subalgebra that is a von Neumann factor assuming that $m = {\rm dim} \mathcal{H}_\partial / {\rm dim} \mathcal{H}_A$ is an integer, guaranteed if all the bond dimensions are the same. To do this, we choose isometries $T^{(a)}$, $a=1 \dots m$ such that $T^{(1)} = T_A$ and the $T^{(a)}$ map into orthogonal subspaces of $\mathcal{H}_\partial$. In this case, 
\[
U_A \equiv (T^{(1)} \dots T^{(m)})~,
\]
defines a unitary from $\mathcal{H}_A \otimes \mathbb{C}^m$ to $\mathcal{H}_\partial$. Using this, we can define an operator map
\[
\varphi(\mathcal{O}_A) = U_A ( \mathcal{O} \otimes \mathbb{1}) U_A^\dagger = \sum_a T^{(a)} {\cal O}_A T^{(a)}{}^\dagger \; .
\]
The image defines a subalgebra of the boundary algebra that is isomorphic to $B(\mathcal{H}_A)$ (and is therefore a factor) and includes the identity of the boundary algebra.

An interesting point here is that in requiring that the boundary algebra is a von Neumann algebra and a factor (as we proposed in section \ref{sec:gen_ent_wedge}), we had to make choices for the isometries $T^{(a)}, a > 1$. This means that the assignment of a boundary subalgebra to a region is non-unique. Instead, we have a whole family of boundary subalgebras, related by the unitary transformations that fix the subspace $T_A \mathcal{H}_A$. This suggests that if there is a connection between BP wedges and algebras in the full gravity setting, it may be more subtle than the simple proposals of section \ref{sec:gen_ent_wedge}, and may also involve the assignment of a family of unitarily equivalent algebras to a BP wedge rather than a single unique algebra. This freedom may be related to the quantum error correcting properties of holographic states.

For the simple tensor networks without bulk legs, we can check whether the proposed algebraic interpretation (\ref{AlgRT}) of generalized entropy gives a sensible result. Here, we can take the algebra $\Omega$ to be the full algebra of operators acting on the boundary Hilbert space, and choose the constant $K_\Omega$ to be the logarithm of the dimension of this algebra (the square of the dimension of the boundary Hilbert space). In this case, we have a pure state on ${\cal H}_A$ created by the tensor network in the region $A$. The boundary state is the image of this state under the isometry $T_A$, and the algebraic entropy of this boundary state for the boundary subalgebra associated with $A$ is equal to the  algebraic entropy for the original state on the algebra of operators on ${\cal H}_A$, which is just $-\log D_A$ since the von Neumann entropy is zero. The index gives $\log D_A^2 - \log D_{\Omega}$, so altogether we have 
\[
S(\omega | \mathcal{A}_A) - \log {\rm Ind}(E_{\Omega \to \mathcal{A}_A}) + K_\Omega = \log D_A = N_A \log D~,
\]
where $N_A$ is the number of bulk legs and $D$ is the bond dimension. This is just what we would expect for the tensor network version of $S_{gen}$, since we don't have an entropy associated with the bulk state and tensor network version of the area term ${\rm Area} / (4G)$ is $\log D$ times the number of legs on the boundary of the region.\footnote{An interesting point is that for a tensor network subsystem that includes a part of the boundary, the $N_A$ includes the boundary legs, so in the gravity case, the quantity we assign to entropy of a generalized entanglement wedge that includes a part of the conformal boundary includes a formally divergent term associated with the area of this part of the conformal boundary. However, for such regions, we should only be considering entropy differences for regions in the same class, and in these differences, we can think of this divergent piece as cancelling along with usual  divergent piece coming from the part of the region boundary that approaches the conformal boundary.}

A more detailed study of the tensor network model will appear in \cite{toappear}.

\subsection{Direct construction of gravitational algebras}

\label{sec:grav_alg}

Another interesting setting in which to explore our proposal is the direct construction of gravitational algebras in semiclassical quantum gravity. Gravitational observables differ fundamentally from those of local quantum field theory. In gauge theories, physical states must satisfy gauge constraints, which obstruct the decomposition of the global Hilbert space into a tensor product of Hilbert spaces associated to local regions. This obstruction is more severe in gravity, where diffeomorphism invariance is the gauge symmetry and acts not on fields over a fixed background but on the spacetime manifold itself \cite{Donnelly:2014fua,Donnelly:2016auv}. As a consequence, local operators are not diffeomorphism-invariant, and the appropriate notion of a subsystem in semiclassical gravity is given by algebras of observables that are suitably dressed to the gauge constraints — for instance, by anchoring to a boundary, attaching gravitational Wilson lines, or  defining observables relationally using matter fields. Various techniques for gravitationally dressing observables and their implications for the corresponding algebras have been studied extensively; see e.g.~\cite{Giddings:2005id,Giddings:2015lla, Donnelly:2015hta,Donnelly:2016rvo, Giddings:2018umg}.

Recent constructions identify von Neumann algebra of observables associated to a spacetime subregion in the limit $G_N \to 0$. In this regime, gravitational backreaction is negligible, and quantum field theory on a fixed background geometry provides the leading approximation, while gravitational corrections are incorporated perturbatively in $G_N$. While the initial constructions of such algebras relied on specific geometries like static patch in dS space \cite{Chandrasekaran:2022cip} or the AdS black hole \cite{Chandrasekaran:2022eqq}, more recent works have argued that the construction can be implemented more generally in semiclassical gravity \cite{Jensen:2023yxy, AliAhmad:2023etg, Klinger:2023tgi, Kudler-Flam:2023qfl, Kudler-Flam:2023hkl, Klinger:2023auu, Faulkner:2024gst, Kudler-Flam:2024psh, AliAhmad:2024eun, Speranza:2025joj} under additional assumptions. 

A key element in the construction is the implementation of gravitational constraints arising from diffeomorphism invariance. Enforcing these constraints modifies the algebraic structure of subregions in a crucial way. A concrete realization of this was provided in \cite{Chandrasekaran:2022cip}, where an observer degree of freedom was introduced within the subregion to serve as a gravitational anchor for diffeomorphism invariant observables. It was then argued that the same mechanism can be used to define gravitational algebras for arbitrary bounded subregions. The resulting algebra, which incorporates a suitable notion of gravitational dressing, is naturally expressed as a modular crossed product algebra. Starting from the type $III_1$ von Neumann algebra associated to local quantum field theory, the corresponding modular crossed product is a type $II$ algebra. Unlike type $III$ algebras, type $II$ von Neumann algebras possess a faithful trace, allowing for a finite, renormalized notion of entropy that has been shown to match the generalized entropy up to a state-independent constant. 

Furthermore, when an appropriate positive energy projection is imposed on the observer sector, the resulting algebra becomes type $II_1$ and admits a maximal entropy state. For unbounded regions extending to infinity, the corresponding algebra is type $II_\infty$. 

Both the BP proposal and the gravitational algebras are supposed to address the same regime of semiclassical gravity. It would therefore be interesting to investigate to what extent our proposal is realized for the assignment of modular crossed product algebras to subregions. In this prescription, one associates to each bounded BP wedge $W$ a type $II_1$ von Neumann algebra $\hat{\mathcal{A}}_W$. In particular, the algebra $\hat{\mathcal{A}}_W$ exhibits a notion of finite entropy, which is directly related to the generalized entropy for a specific class of states. While many properties in our proposal are realized for a local net of QFT algebras, it is by no means clear that they remain true for the corresponding net of gravitational algebras. For example, due to the state-dependence of the gravitational dressing one cannot straightforwardly compare gravitational algebras associated to different BP wedges. Fortunately, the state-dependence of the modular crossed products is rather mild in the sense that crossed product algebras with respect to different states are related by a unitary. This suggests that one should perhaps only expect our proposal to be realized up to unitary equivalence. A more detailed investigation of this will be carried out in forthcoming work.

We would like to stress that there are many assumptions that go into the gravitational algebra construction. Imposing a single boost constraint about the entangling surface already suffices to produce type $II$ gravitational algebras, even for subregions that are not necessarily BP wedges. Yet, one expects that in the full gravitational description an infinite family of such constraints needs to be implemented, one for every compactly supported diffeomorphism \cite{Giddings:2025bkp, Giddings:2025xym}. In the construction of \cite{Jensen:2023yxy}, it is assumed that most of these constraints, coming from diffeomorphisms that act only within the subregion, can be implemented perturbatively through an appropriate  dressing of the operators, in such a way that the algebra and its commutant remain type $III_1$ von Neumann algebras. The explicit boost constraint is special in the sense that it relates the dressing of the subregion to the one of the complementary subregion. More generally, one would need to consider diffeomorphisms that deform the subregion boundaries \cite{Klinger:2026tws}.  This relates to the problem of specifying subregions in a diffeomorphism invariant way. As was suggested in \cite{Jensen:2023yxy}, a useful gauge-fixing condition is to impose the extremization of a suitable geometric functional. This suggests that a construction of the full gravitational algebra, which takes into account all the gravitational constraints, requires the underlying subregion to satisfy conditions that are analogous to the ones defining a BP wedge. 

\subsection{Further directions}

\label{sec:fut_dir}

There are various other directions to explore beyond those described in the previous subsections. First, in this paper we have restricted to the simple case of BP wedges on a slice fixed by time-reflection symmetry. But these wedges have been defined more generally for arbitrary spacetimes \cite{Bousso:2022hlz,Bousso:2023sya,Bousso:2025fgg} so it is natural to explore how the proposed connection with algebras might extend to this more general set of wedges.

\paragraph{Background independence of the algebraic structure?}
If the proposed map from generalized entanglement wedges and algebras exists in some form, it will be interesting to understand the extent to which the resulting algebraic structure is background-independent. At first it may appear that the entire discussion is very background dependent, since BP wedges are defined starting from from a particular fixed geometry. However, it is possible that for two different geometries, we could have a one-to-one corresponding between the generalized entanglement wedges. This is the case for ordinary entanglement wedges for asymptotically AdS spacetimes. If this holds for more general wedges, it might be the case that different geometries (within some equivalence class) have generalized entanglement wedges that correspond to the same underlying set of algebras, with the difference between the geometries coming from the different states on these algebras. See \cite{Witten:2025xuc, Liu:2025krl} for some closely related comments. Some evidence that this might be true comes from the observation above that all BP wedges in AdS${}_3$ can be described via joins of intersections of usual entanglement wedges. Thus, starting from a BP wedge in AdS${}_3$, we can express it as a join of intersections of usual entanglement wedges corresponding to some specific set of boundary regions. For any other asymptotically AdS${}_3$ geometry, we can then define a corresponding BP wedge by taking the join of the intersection of the entanglement wedges corresponding to the same boundary regions. This defines a map from BP wedges in AdS${}_3$ to BP wedges in the new asymptotically AdS geometry. If the new geometry has no other additional BP wedges, then this would be the desired 1-1 correspondence. It would be interesting to understand more generally whether deformations of a spacetime preserve the structure of entanglement wedges; a positive result would provide additional evidence for a fixed algebraic structure in the underlying theory. 

\paragraph{Algebraic origin for gravitational equations?}

A nice application of the usual RT formula in holography is the demonstration that bulk gravitational equations  (at least perturbatively)  have a quantum information theoretic origin, e.g. from positivity of relative entropy \cite{Faulkner:2013ica,Lashkari:2013koa,Swingle:2014uza,Lewkowycz:2018sgn}. The simplest version of these derivations start by considering a particular boundary subsystem, varying the state, and understanding the gravitational implications of positivity of relative entropy between the original state and the perturbed state. Assuming that we have various spacetimes that share the same algebraic structure as described in the previous paragraph, it would be interesting to consider the gravitational implications of algebraic relative entropy positivity and monotonicity for algebras associated with generalized entanglement wedges. As noted already in \cite{Jensen:2023yxy} this may provide an underlying framework for a derivation of the Einstein's equations similar to Jacobson's entanglement equilibrium derivation  \cite{Jacobson_2016}.

\paragraph{Modular Berry phase}
Given the assignment of algebras to BP wedges, it is natural to ask whether the resulting ``net'' of algebras carries any non-trivial global information. A useful probe for this is the modular Berry phase \cite{Czech:2019vih}, a geometric phase associated with a one-parameter family of density matrices. This quantity has recently been defined abstractly in the context of arbitrary von Neumann algebras \cite{Czech:2023zmq, deBoer:2025rxx} using modular theory, making it a particularly suitable observable in the present setting. Exploring the modular Berry phase for our BP wedge algebras may reveal new aspects of their role in the holographic correspondence. 

\paragraph{Gauge-invariant subregions}

Defining fully gauge-invariant subregions and observable algebras in gravity is a non-trivial problem, which has motivated various mechanisms for dressing an observable. These include relational observables, gravitational dressing \cite{Donnelly:2015hta, Giddings:2005id, Giddings:2025xym}, dressing with respect to a wave function \cite{Jensen:2024dnl}.
As noted in section \ref{sec:grav_alg}, a complete implementation of gauge constraints will be crucial in understanding how an extremality condition could emerge while constructing type~$II$ gravitational algebras \cite{Giddings:2025bkp}.
While simultaneously imposing all diffeomorphism constraints remains challenging, some insights can be drawn from discrete models such as gauged tensor networks \cite{Dong:2023,Akers_2024,Qi_2022,Balasubramanian:2025}. 

\paragraph{Tensor networks and quantum error correction}
In section \ref{sec:RTN}, we employ random tensor networks to investigate the validity of our proposal, as they provide a toy model which displays a clean emergence of RT-like extremality conditions in certain regimes. It would be interesting to consider more sophisticated tensor network models that accommodate infinite-dimensional algebras, like the ones studied in \cite{Kang:2019dfi, Gesteau:2020hoz, Chemissany:2025vye}.
Another related work is \cite{Soni:2023} which provides a route for turning type $I$ algebras with a nontrivial centre—interpreted as an area operator in holographic codes \cite{harlow_ryutakayanagi_2017,Kamal:2019skn}—into something resembling a crossed product structure. Since gauging random tensor networks also introduces a nontrivial area operator by modifying the bulk algebras \cite{Dong:2023,Akers_2024,Qi_2022,Balasubramanian:2025}, it would be natural to investigate the role of gauge constraints and extremality conditions in these toy models. In addition, tensor networks realize certain quantum error correcting features that one expects from holography. Given our proposal, it would be interesting to study the quantum error correcting properties of the generalized entanglement wedge algebras using the algebraic framework developed in, e.g., \cite{Kang:2018xqy, Kang:2019dfi, Faulkner:2020hzi, Faulkner:2020iou, Faulkner:2020kit, Furuya:2020tzv, Gesteau:2021jzp, Faulkner:2022ada, vanderHeijden:2024tdk, Crann:2024gkv, AliAhmad:2024saq, AliAhmad:2025oli}. In that context, the existence of a conditional expectation is closely related to a notion of \emph{exact} quantum error correction, while in general a modified algebraic structure resembling \emph{approximate} quantum error correction might be necessary (see e.g., \cite{AliAhmad:2024saq, AliAhmad:2025oli}). We leave an investigation of this to future work.

\section*{Acknowledgements}

We would like to thank Chris Akers, Raphael Bousso, Netta Engelhardt, Tom Faulkner, Elliott Gesteau, Marc Klinger, Nima Lashkari, Samuel Leutheusser, Hong Liu, Kristan Jensen, Pompey Leung and Geoff Penington for valuable discussions. This work is supported in part by the National Science and Engineering Research Council of Canada (NSERC) and the Simons Foundation via a Simons Investigator Award.

\appendix

\section{Algebraic quantum information}
\label{app:algebraicQI}

In this section, we discuss in general how many of the basic quantum information theory concepts used in holography generalize to the algebraic setting where we consider restrictions of states to general subalgebras of the algebra of observables.

This generalization is already essential when considering spatial subsystems of a continuum quantum field theory. In this paper, we will apply it to consider subsystems of a general gravitational theory. But we emphasize that even in the simple setting of quantum mechanics on a finite-dimensional Hilbert space, the algebraic approach provides a more general set of quantum information theoretic quantities than the usual ones that are constructed by assuming a tensor factorization. For example, even the quantum system of a single qubit has non-trivial proper subalgebras to which we can assign entropies that quantify the information in a coarse-grained version of the state.

\subsection{General quantum systems}

A very general notion of a quantum system starts with a net of algebras of observables, that is, a collection $\{\mathcal{A}_i\}$ of unital $C^*$ algebras\footnote{Rough algebra primer: a $*$-algebra is set of operators closed under the operations of addition, multiplication by a complex scalar, multiplication, and adjoint (the $*$ operation). For a $C^*$ algebra, we also have a norm satisfying $||a^* a|| = ||a||^2$ and demand closure under this norm. Unital means that the algebra includes the identity operator.} indexed by some partially ordered set $I$, with inclusion maps (technically, unital injective *-homomorphisms) $\iota_{ij} : \mathcal{A}_i \to \mathcal{A}_j$ for $i \le j$. A state of this quantum system is a set of positive linear maps $\omega_i : \mathcal{A}_i \to \mathbb{C}$ with compatibility conditions $\omega_i = \omega_j \circ \iota_{ij}$ for $i \le j$. 

Often, there is some global algebra $\mathcal{A}_{0}$ containing all the other algebras as subalgebras, and a global state $\omega$ with all the other states obtained by the inclusion maps $\omega_i =  \omega \circ \iota_{i 0}$.\footnote{Such a global algebra can be defined any time the set $I$ is directed, i.e. for every $\{i,j\} \subset I$ we have $k \in I$ with $i \le k$ and $j \le k$. In this case, we can define a global algebra $\mathcal{A}_0$ by taking a completion of the union of all the operators in all the various algebras with equivalence between ${\cal O}_i \in \mathcal{A}_i$ and ${\cal O}_j \in \mathcal{A}_j$ if there is a $k$ with $k \ge i$ and $k \ge j$ and $\iota_{ik} {\cal O}_i = \iota_{jk} {\cal O}_j$.}

Given such a quantum system in a particular state, there are various quantum information theoretic measures that we can use to characterize the subalgebras and their relationship and also the state and its various restrictions to particular algebras.

\subsection{Characterizing subalgebras and their relationships}

Given any algebra $\mathcal{A}_i$ and its corresponding state $\omega_i$, we can always (via the GNS construction) construct a Hilbert space ${\cal H}_i$ with a state $|\Omega_i \rangle$ and a representation $\pi$ of $\mathcal{A}_i$ on ${\cal H}_i$ such that values of all observables in the state $\omega_i$ are obtained by inner products on the Hilbert space:
\[
\omega_i(a) = \langle \Omega_i| \pi(a) |\Omega_i \rangle \qquad \forall a \in \mathcal{A}_i \; .
\]
For physical applications, it is natural to enlarge the algebra $\mathcal{A}_i$ to include all operators ${\cal O}$ acting on ${\cal H}_i$ whose matrix elements are limits of the matrix elements for a sequence of operators in $\mathcal{A}_i$. According to a theorem of von Neumann, this is equivalent to including all bounded operators on ${\cal H}_i$ in the double-commutant of $\mathcal{A}_i$.\footnote{The commutant ${\cal A}_i'$ of ${\cal A}_i$ is the set of operators acting on ${\cal H}_i$ that commute with ${\cal A}_i$. The double commutant ${\cal A}_i''$ is the set of operators that commute with ${\cal A}_i'$.} The resulting algebra is a von Neumann algebra $\hat{\mathcal{A}}_i = \mathcal{A}_i''$. Given a collection of von Neumann algebras $\{\hat{\mathcal{A}}_i\}_{i\in I}$ their union $\bigcup_{i\in I} \hat{\mathcal{A}}_i$ is not automatically a von Neumann algebra. We denote by $\bigvee_{i\in I} \hat{\mathcal{A}}_i$ the smallest von Neumann algebra that contains the union, obtained by taking its double commutant. The intersection of von Neumann algebras does define a von Neumann algebra, denoted by $\bigwedge_{i\in I} \hat{\mathcal{A}}_i$.

A first basic ``information-theoretic'' characterization of our quantum system is the type classification for the various von Neumann algebras $\hat{\mathcal{A}}_i$. Each von Neumann algebra has a {\it center} ${\cal Z}(\hat{\mathcal{A}}_i)$ defined as the set of elements in $\hat{\mathcal{A}}_i$ that commute with all of $\hat{\mathcal{A}}_i$. If the center includes only multiples of the identity operator, the von Neumann algebra is known as a {\it factor}. More generally, the center is an Abelian von Neumann algebra, and any such algebra is isomorphic to the space $L^\infty(X,\mu)$ of functions on some space $X$ with measure $\mu$. In this general case, the algebra can be written as a direct integral (a generalization of direct sum) of von Neumann factors, one for each point in $X$. When $X$ is discrete, the algebra is then simply a direct sum of factors. 

There is a standard classification for the individual factors. To understand the structure of a factor $\mathcal{M}$, it is very useful to consider the set of projections, operators $p$ in $\mathcal{M}$ with $p^2 = p = p^*$. These correspond to yes/no observables in quantum mechanics; the expectation value $\omega(p)$ tells us the probability that our state will be found to have a particular property in a measurement. The set of projections generate the full algebra. Two projections $p$ and $q$ are considered to be equivalent if there is an operator $v$ in $\mathcal{M}$ (a ``partial isometry'') such that $p = v^* v$ and $q = v v^*$. In finite dimensions, we can alternatively require that there is a unitary $u$ with $q = u p u^*$, so equivalence means intuitively that we are projecting to equivalent subspaces. 

There is a natural ordering among projections that we can understand as characterizing the ``size'' of the space that we are projecting to. Given a projection $p$, a sub-projection $p_1$ is a projection satisfying $p_1 p = p p_1 = p_1$. We have that a projection  $p_1$ is a subprojection of $p$ if and only if $p_1 \mathcal{M} p_1 \subset p \mathcal{M} p$. If $q$ is equivalent to a subprojection of $p$, we write $q \precsim p$. It turns out that {\it any} two projections $p$ and $q$ in a factor can be ordered in this way: either $p \precsim q$ or $q \precsim p$. Further, there is a {\it dimension} function $d(p)$ on projections, unique up to a multiplicative rescaling, with the properties that
\[
d(p) \le d(q) \iff p \precsim q \qquad d(p) = d(q) \iff p \sim q \qquad d(0) = 0 \qquad pq = 0 \implies d(p \oplus q) = d(p) + d(q) 
\]
The classification of von Neumann factors is captured neatly via the range of this dimension function (allowing for a rescaling):
\begin{itemize}
    \item {\bf Type $I_n$}: $d(p) \in \{0,1,2,\dots,n\}$. These factors are equivalent to the full set of operators acting on a Hilbert space of dimension $n$. 
    \item {\bf Type $I_\infty$}:  $d(p) \in \{0,1,2,\dots,\infty\}$. These factors are equivalent to the full set of bounded operators acting on an infinite dimensional Hilbert space. 
    \item {\bf Type $II_1$}: $d(p) \in [0,1]$. Here, we have a finite trace $\tau: \mathcal{M} \to \mathbb{C}$ (i.e., a functional with the tracial property $\tau(ab)=\tau(ba)$) which is normalized such that $\tau(1) = 1$.
    \item {\bf Type $II_\infty$}: $d(p) \in [0,\infty]$. Here, we have a trace that is finite on a dense subset of the operators (called ``semi-finite'').
    \item {\bf Type $III$}: $d(p) \in \{0,\infty\}$. In this case, there is no non-zero semi-finite trace. Roughly, any projection reduces the dimension by an infinite amount. Example: algebras of operators associated to QFT spatial subsystems with a non-trivial boundary.
\end{itemize}

\subsection{Entropy and relative entropy}

For standard quantum subsystems with operator algebra $\mathcal{A}=M_n(\mathbb{C})$, we can use entropy to quantify the amount of information and/or uncertainty that is present in the reduced state on a subsystem. We typically construct the reduced density operator $\rho$ for the subsystem in terms of the matrix trace $\tr$, and then define the von Neumann entropy
\[
S = - \tr(\rho \log \rho) \; .
\]
This number quantifies our ignorance about the state, ranging from 0 when the subsystem is in a definite pure state to $\log D$ (where $D$ is the subsystem dimension) for the maximally mixed state where we have no information. We can say that 
\[
\log D - S 
\]
is the number of bits of information that we learn about the state by knowing the values of all possible observables associated with the algebra of operators acting on the subsystem. In the infinite dimensional case, we may have certain states with finite entropy and others where the entropy diverges.

We can often extend this to the case of a more general subalgebra of observables. In cases where we have an algebra $\mathcal{M}$ with a faithful, normal, semi-finite\footnote{Since we are generally working with infinite dimensionsal algebras, it is necessary to keep track of some technical requirements. We say that weight is \emph{faithful} if it assigns a strictly positive value to each non-zero positive operator. It is \emph{normal} if it is continuous in some appropriate sense (namely in the ultra-weak topology). A weight is \emph{semi-finite} if it is finite on a subset of operators that is dense in the full algebra.} trace $\tau$ (i.e., not type $III$), then every normal state $\omega$ on $\mathcal{M}$ may be written as
\[
\omega(a) = \tau(\rho a)~, 
\]
for some unique positive operator $\rho \in \mathcal{M}$ with $\tau(\rho) = 1$. In this case, we can define an entropy for the state $\omega$ as
\[ \label{eq:algebraicentropy}
S(\omega|\mathcal{M}) = - \tau(\rho \log \rho) \; .
\]
This entropy is always finite for the type $I_n$ case but otherwise can be finite or infinite depending on the state.

Even when the entropy of a state is infinite or cannot be defined, it is still possible to compare the information content of two sufficiently nearby states using relative entropy. Generalizing the usual definition for subalgebras, we have in the tracial case for faithful states $\omega, \sigma$ with densities $\rho_\omega,\rho_\sigma$,
\[
S(\omega\Vert\sigma)\ =\ \tau\big(\rho_\omega(\log\rho_\omega-\log\rho_\sigma)\big)~.
\]
In the type $III$ case, it is still possible to define relative entropy between states using Tomita-Takesaki theory \cite{Araki:1976zv, Araki:1977zsq}. In this type $III$ case, the relative entropy is always finite between faithful states. For the other cases, the relative entropy is finite if and only if the support of $\omega$ is contained in the support of $\sigma$.

When finite, the relative entropy is always a positive quantity. In the tracial case, we can take $\sigma$ to be the tracial state such that the second term vanishes and we have
\[
S(\omega || \tau) = -S(\omega|\mathcal{M}) \; .
\]
This shows that the entropy defined above is always negative. We showed in \cref{sec:alg_RT} that for a standard algebra of observables on a finite dimensional Hilbert space of dimension $D$, the algebraically defined entropy is related to the usual von Neumann entropy by $S(\omega) = S_{vN} - \log D$.

\subsection{Coarse-graining and conditional expectations}
\label{sec:cond_expectation}

In the case case where an algebra $\mathcal{M}$ has a subalgebra $\mathcal{N}$, there are various ways that we can characterize the relative size of the subalgebra and also the relative amount of information contained in the state $\omega_\mathcal{N}$ obtained by restricting the a state $\omega$ on $\mathcal{M}$ to the subalgebra.

A key concept is the idea of a {\it conditional expectation}. This is a linear map $E:\mathcal{M} \to \mathcal{N}$ with the properties that:
\begin{eqnarray*}
   (i)&&\; E(xx^*) \geq 0~, \; \forall \; x \in \mathcal{M}~, \cr
   (ii)&& \; E(y) = y~, \; \forall y\in \mathcal{N}~, \cr
   (iii)&& \; E(x y) = E(x) y~, \; \forall x \in \mathcal{M}~, \forall y\in \mathcal{N}~.
\end{eqnarray*}
In addition, one often requires that the conditional expectation is compatible with some chosen faithful normal state\footnote{More generally, $\omega$ can be a faithful semi-finite normal weight.} $\omega$ in the sense that 
\begin{eqnarray*}
   (iv)&& \; (\omega\circ E)(x) = \omega(x)~, \; \forall x \in \mathcal{M}~.
\end{eqnarray*}
The map $E$ is then automatically faithful and normal as well. See for example \cite{1993QuantumEA} for more details. The conditional expectation ``coarse-grains'' operators in $\mathcal{M}$ to operators in the subalgebra $\mathcal{N}$ in a way that preserves expectation values in the state $\omega$. 
For algebras with a trace, we will typically be interested in the case where $\omega$ is the tracial state $\tau$. When the trace is finite, we have a unique trace-preserving conditional expectation that is simply projection to $\mathcal{N}$ with respect to the inner product $\langle b , a \rangle = \tau(b^* a)$. 

The conditional expectation provides us with an algebraic analog of the familiar partial trace operation that gives us the state of a subsystem. Given the density operator $\rho$ associated with a state $\omega$, the state $\rho_\mathcal{N} = E(\rho)$ gives the density operator in $\mathcal{N}$ corresponding to the state $\omega$ restricted to $\mathcal{N}$: 
\[
\omega(b) = \tau(\rho b) = \tau(E(\rho b)) = \tau(E(\rho) b) = \tau(\rho_\mathcal{N} b) \;, \quad \forall b \in \mathcal{N}~.
\]
We can also use the conditional expectation to produce a coarse-grained state 
\[
\omega \circ E 
\]
on the original algebra $\mathcal{M}$. This preserves the information about all observables the subalgebra $\mathcal{N}$ but discards as much other information as possible. It has the properties that it 
\begin{itemize}
    \item is the unique state of maximal entropy among all states that agree with $\omega$ on $\mathcal{N}$.
    \item is the unique minimizer of relative entropy $S(\omega || \sigma)$ among states $\sigma$ with densities in $\mathcal{N}$.
\end{itemize}

The relative entropy $S(\omega || \omega \circ E)$ quantifies how much information has been lost in the coarse-graining to $\mathcal{N}$. When the individual entropies of the state $\omega$ and its restriction to $\mathcal{N}$ are finite, we have
\begin{equation}
    \label{Sdiff3}
S(\omega || \omega \circ E) = S(\omega| \mathcal{N}) - S(\omega| \mathcal{M})~,
\end{equation}
where the right-hand side is a difference of algebraic entropies defined in \eqref{eq:algebraicentropy}, i.e., it is the increase in entropy under coarse-graining. One can show that this relative entropy obeys 
\[
0 \le S(\omega || \omega \circ E) \le \log {\rm Ind}(E) \; ,
\]
where ${\rm Ind}(E)$, to be defined presently, is a quantity called the index that measures the relative size of $\mathcal{M}$ and $\mathcal{N}$.

\subsubsection{The index}

\label{sec:index}

In finite dimensions, an algebra and a subalgebra each have a certain dimension, and we can consider the ratio $\dim(\mathcal{M})/\dim(\mathcal{N})$ of these dimensions as a measure of how large $\mathcal{M}$ is relative to $\mathcal{N}$. For more general cases, this ratio generalizes to the notion of an {\it index}. 

Let $\mathcal N\subset\mathcal M$ be an inclusion of von Neumann factors and let $E:\mathcal M\to\mathcal N$ be a faithful normal conditional expectation. Associated with $E$ is a canonical map\footnote{Technically, $E^{-1}$ is an operator-valued weight defined in terms of $E$ using Connes' spatial theory \cite{CONNES1980153}.} $E^{-1}:\mathcal N'\to\mathcal M'$ on the commutant algebras, and the \emph{Jones–Kosaki index} is defined by \cite{kosaki1986}
\[
{\rm Ind}(E):=E^{-1}(1)~.
\]
For factors, the index is a number in $[1,\infty]$ with ${\rm Ind}(E) = 1$ if and only if $\mathcal{N} = \mathcal{M}$. When $\mathcal{M}$ and $\mathcal{N}$ are each finite factors (i.e. type $I_n$ or $II_1$) and $E$ is the unique trace-preserving conditional expectation, this index agrees with the independently defined {\it Jones index} \cite{jones1983index}, denoted by $[\mathcal{M}:\mathcal{N}]$, which has been shown to take values in $\{4 \cos^2 {\pi \over n}\, | \, n = 3,4,\dots \} \cup [4,\infty]$. The Jones index is defined as $[\mathcal{M}:\mathcal{N}]=(\tau_{\mathcal{N'}}(e))^{-1}$, where $e$ is the projector that realizes the conditional expectation $E$ on the level of the GNS Hilbert spaces. 

The above definition is closely related to the \emph{Pimsner–Popa index} \cite{Pimsner1986}, defined as the infimum of $\lambda^{-1}$ over all constants $\lambda > 0$ such that 
\[
E(x^* x) \ge \lambda x^* x~, \; \; \;  \forall x \in \mathcal{M}~.
\]
For infinite factors these notions agree under the usual finite-index assumptions, while in finite-dimensional type $I$ algebras the Pimsner–Popa index can differ from the Jones index (see e.g. \cite{Longo:2022lod}).

In particular, for an inclusion of the form
\[
\mathcal{N}\equiv M_n(\mathbb C)\otimes 1_m
\subset
\mathcal{M}\equiv M_n(\mathbb C)\otimes M_m(\mathbb C)~,
\]
and $E=\id_{M_n}\otimes\tau_m$ one finds that
\[
{\rm Ind}(E)
=
\frac{\dim_{\mathbb C}(\mathcal M)}
{\dim_{\mathbb C}(\mathcal N)} 
= m^2~.
\]
Thus, for finite-dimensional factors, the Jones–Kosaki index of the trace-preserving expectation is simply the ratio of the dimensions of the two matrix algebras, or equivalently the square of the ratio of the dimensions of their defining Hilbert spaces.

There are a couple of useful properties of the index that we will require. First, if we have inclusions $\mathcal{M} \subset \mathcal{N} \subset \mathcal{P}$ of von Neumann algebras each with a faithful normal tracial state 
and faithful normal trace-preserving conditional expectations $E_{\mathcal{N} \to \mathcal{M}}$, $E_{\mathcal{P} \to \mathcal{N}}$, and $E_{\mathcal{P} \to \mathcal{M}} \equiv  E_{\mathcal{N} \to \mathcal{M}}\circ E_{\mathcal{P} \to \mathcal{N}}$ ,  we have in general that if the indices are finite,
\begin{equation}
\label{IndId}
{\rm Ind} E_{\mathcal{P} \to \mathcal{M}} \le {\rm Ind} E_{\mathcal{P} \to \mathcal{N}} \; {\rm Ind} E_{\mathcal{N} \to \mathcal{M}}~,
\end{equation}
with equality when the algebras are factors and the expectations are of {\it minimal index}, that is, when there is no other faithful normal conditional expectation with a smaller index \cite{longo1989index}. For $II_1$ factors, the unique trace-preserving conditional expectations have minimal index, so the equality is satisfied. 

Focusing on the type $II_1$ factor case where we have equality, suppose now that we have subalgebras $\mathcal{B}$ and $\mathcal{C}$ with $\mathcal{A} = \mathcal{B} \vee \mathcal{C}$ and $\mathcal{D} = \mathcal{B} \wedge \mathcal{C}$. In this case, the trace-preserving conditional expectations have minimal index, and we have
\[
{\rm Ind} E_{\mathcal{A} \to \mathcal{D}} =  {\rm Ind} E_{\mathcal{A} \to \mathcal{B}} \; {\rm Ind} E_{\mathcal{B} \to \mathcal{D}} = {\rm Ind} E_{\mathcal{A} \to \mathcal{C}} \; {\rm Ind} E_{\mathcal{C} \to \mathcal{D}} \; .
\]
If in addition the {\it commuting square condition }  
\[
E_{\mathcal{A} \to \mathcal{B}}\circ E_{\mathcal{A} \to \mathcal{C}} = E_{\mathcal{A} \to \mathcal{C}} \circ E_{\mathcal{A} \to \mathcal{B}}~,
\]
is satisfied, we have the following index inequality \cite{Popa1989RelativeDT}:
\begin{equation}\label{eq:indexineq}
{\rm Ind} E_{\mathcal{B} \to \mathcal{D}} \le {\rm Ind} E_{\mathcal{A} \to \mathcal{C}}~, \qquad \qquad {\rm Ind} E_{\mathcal{C} \to \mathcal{D}} \le {\rm Ind} E_{\mathcal{A} \to \mathcal{B}}~.
\end{equation}
We can formulate an analogous notion for the commutant algebras $\mathcal{B}'$ and $\mathcal{C}'$ with $\mathcal{A}'=\mathcal{B}'\wedge \mathcal{C}'$ and $\mathcal{D}'=\mathcal{B}'\vee \mathcal{C}'$. We say that the \emph{co-commuting square condition} is satisfied if
\[
E_{\mathcal{D}' \to \mathcal{B}'}\circ E_{\mathcal{D}' \to \mathcal{C}'} = E_{\mathcal{D}' \to \mathcal{C}'} \circ E_{\mathcal{D}' \to \mathcal{B}'}~.
\]
A commuting square of algebras satisfying the co-commutativity condition is called \emph{non-degenerate}, and there are several equivalent characterizations. One such characterization is that for a non-degenerate commuting square we have equality in \eqref{eq:indexineq}. Combining results, we have under the commuting square condition
\[
{\rm Ind} E_{\mathcal{A} \to \mathcal{D}} \le {\rm Ind} E_{\mathcal{A} \to \mathcal{C}} \; {\rm Ind} E_{\mathcal{A} \to \mathcal{B}} \;,
\]
with equality if the commuting square is non-degenerate. Some of these relations are discussed in \cite{jones1997introduction, Takashi, sano1996commuting}.

\subsection{Data processing and algebraic strong subadditivity}

\label{sec:alg_SSA}

The relative entropy between states $\omega$ and $\psi$ on the same algebra $\mathcal{M}$ can be understood as a measure of the distinguishability between the states. A key feature of this is that if we perform any coarse graining, the distinguishability never increases. This stems from:
\paragraph{\bf The Data Processing Inequality}
{\it For von Neumann algebras $\mathcal{M}$ and $\mathcal{N}$, any normal states $\omega$ and $\psi$ on $\mathcal{N}$ and any normal, unital, completely positive map $T: \mathcal{M} \to \mathcal{N}$, 
    \[
    S_\mathcal{N}(\omega || \psi) \ge S_{\mathcal{M}}(\omega \circ T || \psi \circ T) \; .
    \]}
In particular, this applies where $T$ is any conditional  expectation.

It will also be useful to note a certain ``Pythagorean'' identity \cite{petz1991certain} for relative entropies:
\paragraph{\bf The Pythagorean Identity}
{\it For a von Neumann algebras $\mathcal{M}$ with state $\omega$, a von Neumann subalgebra $\mathcal{N}$ with state $\phi$, and a faithful, normal conditional expectation $E: \mathcal{M} \to \mathcal{N}$, we have
\[
S_\mathcal{M}(\omega || \phi \circ E) = S_{\mathcal{M}}(\omega || \omega|_\mathcal{N} \circ E) + S_{\mathcal{N}}(\omega|_\mathcal{N} || \phi) \; .
\]}
Here, relative entropy is playing the role of distance squared in the geometrical analogy.

Now, consider the situation where we have a faithful normal state $\omega$ on some algebra $\mathcal{M}$ and two different subalgebras $\mathcal{A}$ and $\mathcal{B}$ with conditional expectations $E_\mathcal{A}:\mathcal{M} \to \mathcal{A}$ and $E_\mathcal{B}:\mathcal{M} \to \mathcal{B}$. When these conditional expectations commute with each other, 
\[
E_\mathcal{A} \circ E_\mathcal{B} = E_\mathcal{B} \circ E_\mathcal{A} \; ,
\]
the composition defines a conditional expectation $E_{\mathcal{A} \wedge \mathcal{B}}$ to the intersection algebra $\mathcal{A} \wedge \mathcal{B}$. We say that $E_\mathcal{A}$ and  $E_\mathcal{B}$ satisfy the {\it commuting square condition}.

When this holds, we have a version of strong subadditivity. We can obtain this from the following result, proved in \cite{petz1991certain} using the Pythagorean identity and the data processing inequality:
\paragraph{Petz relative entropy inequality}{\it Suppose $\mathcal{M}$ is a von Neumann algebra with separating state $\phi$ and subalgebras $\mathcal{B} $ and $\mathcal{C}$. Let $E_{\mathcal{B}}: \mathcal{M} \to \mathcal{B}$ be a $\phi$-preserving conditional expectation, and suppose that $\mathcal{D} = E_{\mathcal{B}}(\mathcal{C})$ is a von Neumann subalgebra. Then for every state $\omega$, we have
\[
S(\omega || \phi) + S(\omega_\mathcal{D}|| \phi_\mathcal{D}) \ge  S(\omega_\mathcal{B}|| \phi_\mathcal{B}) + S(\omega_\mathcal{C}|| \phi_\mathcal{C}) \; .
\]}
where for example $\omega_\mathcal{B} = \omega \circ E_{\mathcal{B}}$
This applies in particular when we also have $E_{\mathcal{C}} : \mathcal{M} \to \mathcal{C}$ commuting with $E_{\mathcal{B}}$, in which case $\mathcal{D} = \mathcal{B} \wedge \mathcal{C}$ \cite{jones1997introduction}.

We can relate this to the usual strong subadditivity as follows. First, if $\mathcal{M}$ has a tracial state $\tau$, take $\phi = \tau$. Since all the conditional expectations are trace preserving, $\phi_\mathcal{D}=\phi_\mathcal{B}=\phi_\mathcal{C}=\tau$. Given any state $\omega$ we then have $S(\omega_\mathcal{B}||\tau) = - S(\omega|\mathcal{B})$ and similar formulas hold for the subalgebras $\mathcal{C}, \mathcal{D}$. This straightforwardly gives us the following result:

\paragraph{Strong subadditivity}{\it Suppose $\mathcal{M}$ is a von Neumann algebra with a faithful normal finite trace $\tau$ and subalgebras $\mathcal{B}$ and $\mathcal{C}$ with trace-preserving conditional expectations $E_{\mathcal{B}}$ and $E_{\mathcal{C}}$ satisfying the commuting square condition. Then for any state $\omega$
\[
S(\omega|\mathcal{B}) + S(\omega| \mathcal{C}) \ge  S(\omega | \mathcal{M}) + S(\omega | \mathcal{B} \wedge \mathcal{C})   \; .
\]}
In particular, this holds for $\mathcal{M} = \mathcal{B} \vee \mathcal{C}$.

An alternative relative entropy identity that also gives rise to strong subadditivity is
\paragraph{Relative entropy superadditivity}{\it Let $\mathcal{M}$ be a von Neumann algebra with normal state $\phi$ and let $\mathcal{B}$ and $\mathcal{C}$ be subalgebras with commuting faithful normal conditional expectations $E_{\mathcal{B}}$ and $E_{\mathcal{C}}$. Let $\mathcal{N} = \mathcal{B} \wedge \mathcal{C}$. Then 
\[
S(\omega || \omega \circ E_{\mathcal{N}}) \le S(\omega || \omega \circ E_{\mathcal{B}}) + S(\omega || \omega \circ E_C) \; .
\]}
To see this, let $\phi:=\omega\circ E_{\mathcal{N}}$. 
Since $\mathcal{N}\subseteq \mathcal{B},\mathcal{C}$ we have $E_{\mathcal{N}}\!\circ E_\mathcal{B}=E_{\mathcal{N}}$ and $E_{\mathcal{N}}\!\circ E_\mathcal{C}=E_{\mathcal{N}}$, hence both $E_\mathcal{B}$ and $E_\mathcal{C}$ are $\phi$-preserving. The Pythagorean identity (with $E=E_\mathcal{B}$ and reference state $\phi|_{\mathcal{B}}$) gives
\begin{equation}\label{eq:pyth}
S_{\mathcal{M}}(\omega\|\phi)
= S_{\mathcal{M}}(\omega\| \omega\!\circ E_\mathcal{B})
  + S_{\mathcal{B}}\!\big(\,\omega|_{\mathcal{B}} \,\big\|\, \phi|_{\mathcal{B}}\,\big).
\end{equation}
Next, apply data processing for $T=E_\mathcal{B}$ to the pair $(\omega,\ \omega\!\circ E_\mathcal{C})$:
\begin{equation}\label{eq:dpi}
S_{\mathcal{M}}(\omega\| \omega\!\circ E_\mathcal{C})
\ \ge\
S_{\mathcal{B}}\!\big(\,\omega\!\circ E_\mathcal{B}\ \big\|\ (\omega\!\circ E_\mathcal{C})\!\circ E_\mathcal{B}\big)
\ =\
S_{\mathcal{B}}\!\big(\,\omega|_{\mathcal{B}} \,\big\|\, \phi|_{\mathcal{B}}\,\big),
\end{equation}
where we used $(\omega\!\circ E_\mathcal{C})\!\circ E_\mathcal{B}
=\omega\!\circ(E_\mathcal{C}\circ E_\mathcal{B})
=\omega\!\circ E_{\mathcal{N}}=\phi$.

Combining \eqref{eq:pyth} and \eqref{eq:dpi} yields the result. This reduces to the algebraic strong subadditivity result above in the finite entropy case when (\ref{Sdiff3}) applies.

\subsection{Example: tensor factor subalgebras}

To complete this section, we record here what some of the algebraic results reduce to in the familiar situation where we have a quantum system with a tensor product Hilbert space ${\cal H} = {\cal H}_B \otimes {\cal H}_C$ and $\mathcal{A}$ and $\mathcal{B}$ are taken to be the algebra of operators on ${\cal H}$ and the subalgebra of operators of the form ${\cal O}_B \otimes \mathbb{I}_C$.

In this case, using the various normalizations defined above, we have the following:
\[
\begin{array}{ll} \tau & \qquad {1 \over D_{B} D_C} {\rm Tr} \cr
\rho_{\omega} & \qquad D_{B} D_C \rho \cr
E_B(\rho_{\omega}) & \qquad D_B \rho_B \otimes \mathbb{I}_C \cr
S(\omega|\mathcal{B}) & \qquad S_B - \log D_B \cr
S(\omega||\omega \circ E_{\mathcal{B}}) & \qquad [S_B - S_{BC} + \log D_C]
\end{array}
\]

\bibliographystyle{JHEP}
\bibliography{references}

\end{document}